# Energy Harvesting in High Altitude Platform Station Enabled Sensor Networks

Melek Tuylu, *Graduate Student Member, IEEE,* Eylem Erdogan, *Senior Member, IEEE,*

*Abstract*— High altitude platform station (HAPS) systems are becoming crucial facilitators for future wireless communication networks, enhancing connectivity across all vertical communication layers, including small Internet of Things (IoT) sensors and devices, terrestrial users, and aerial devices. In the context of the widely recognized vertical heterogeneous network (VHetNet) architecture, HAPS systems can provide service to both aerial and ground users. However, integrating HAPS systems as a core element in the VHetNet architecture presents a considerable energy challenge, marking a prominent constraint for their operation. Driven by this challenge, we introduce an energy harvesting (EH) strategy tailored for HAPS systems, enabling a HAPS system to gather energy from another HAPS system, which is not constrained by energy limitations. To assess the performance capabilities of the proposed model, we derive outage probability (OP), ergodic capacity (EC) and verify them by using Monte Carlo (MC) simulations. Moreover, we explore the system in terms of throughput. The findings reveal that harnessing full potential of EH stands as a viable approach to meet the energy demands of HAPS systems.

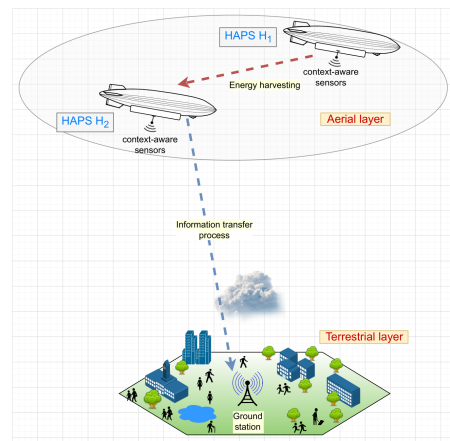

## I. INTRODUCTION

BY 2024, over 50 billion small sensors and Internet of Things (IoT) devices had been connected to the internet and the utilization of small sensors in terrestrial, aerial, and space networks is growing day by day [1]. The integration of small sensors and IoT devices into these networks, as part of the vertical heterogeneous network (VHetNet) architecture, is extending their applications into new domains. Unmanned aerial vehicles (UAVs) for instance, are employed with small sensors, and utilized for collecting environmental data, facilitating precision farming by monitoring soil and crops, and contributing to search and rescue missions with thermal imaging technology [2]. In contrast, high altitude platform station (HAPS) systems utilize IoT devices and small sensors to improve telecommunications, offering connectivity in remote areas, monitoring climate and atmospheric conditions for scientific studies, and bolstering border security with ongoing surveillance. This development represents a significant evolution in the IoT field, expanding the scope of data gathering and connectivity to new levels and delivering cutting-edge solutions in numerous industries.

In the VHetNet architecture, a HAPS system can be viewed as a complementary platform for terrestrial networks to fulfill the growing dynamic capacity need and to create future-proof wireless networks [3], [4]. A HAPS network operates in the stratosphere, where it can provide uplink and downlink communication services [5]. The feasibility of HAPS communication systems has been enhanced in recent times due to advancements in composite materials, computer technology, navigation systems, low-speed high-altitude aerodynamics, sensor devices, and propulsion systems such as internal combustion engines and solar power. In this manner, lighter-than-air vehicles with propulsion and steering systems are one of the well-accepted HAPS designs. Airships, in contrast to typical heavier-than-air vehicles like fixed-wing airplanes and helicopters, generate lift by using a light lifting gas to float through the air. This unique ability can allow them to fly for lengthy periods of time, employ a number of IoT devices and small sensors, carry a heavy burden for their size, and use little fuel [6]. Airships are more cost-effective to operate than traditional aircraft because of their ability to hover for extended periods of time on a single tank of fuel [7]. There is also no need for lengthy runways for boarding.

HAPS systems rely on solar energy for communication needs, and because of this, they experience energy constraints. One of the technologies that can be used to overcome this drawback is energy harvesting (EH). EH is the practice of generating electricity from unusable sources, such as heat, sound, and radio frequency (RF) signals, in order to meet those

Melek Tuylu is with the Faculty of Electrical and Electronics Engineering, Istanbul Technical University 34469 Istanbul, Turkey, e-mail: (tuylu21@itu.edu.tr).

Eylem Erdogan is with the Izmir Institute of Technology, Urla, Izmir, Turkey, e-mail: (eylemerdogan@iyte.edu.tr).

operational needs. Among these sources, EH from RF energy constitutes an important section of the electromagnetic spectrum, and it is frequently preferred in wireless communication systems [8]. In EH, power-constrained receivers process the received RF signals to extract and accumulate usable energy [9]. This harvested energy can then be used to recharge IoT sensors or devices integrated into HAPS systems through the EH approach [10].

**Related Works:** In the vast body of literature, EH has been studied mainly in terrestrial networks. In addition to the studies focused on terrestrial networks, Ghosh et al. [11] have conducted a comprehensive outage analysis in SWIPT-enabled cooperative amplify-and-forward/decode-and-forward (AF/DF) relay systems, highlighting the importance of EH from primary user signals to enhance both energy efficiency and communication reliability. In our study, a linear EH model was chosen due to its simplicity and efficiency, particularly for HAPS where system complexity and overhead are critical factors. While non-linear EH models, such as those discussed in the literature [12] and [13], offer more detailed representations of energy conversion dynamics, they introduce significant computational complexity and are highly sensitive to low-input power conditions. For instance, Alevizos and Bletsas [12] have demonstrated that non-linear models accurately capture efficiency variations at low power levels, but these models require complex parameter fitting and show notable deviations from practical harvesting behavior under real-world conditions. Similarly, Boshkovska et al. [13] highlighted that non-linear models are suitable for scenarios with controlled and predictable input power. However, for the purpose of our system, which operates in diverse and dynamic environments, the linear model provides a more tractable and robust framework without compromising overall system performance.

To further enhance EH efficiency in wireless systems, various methods have been explored. In addition to time switching (TS) EH, the power splitting (PS) technique has also been widely studied for EH in wireless communication systems. For instance, Varshney [14] demonstrated the efficiency of PS in multi-antenna relay systems, showing how it can be used to maximize EH and information transfer. Furthermore, hybrid approaches, such as the combination of PS and TS methods, have been proposed to offer flexibility and enhanced performance under different channel conditions [15]. While PS allows simultaneous EH and decoding, the TS method provides a simpler architecture with lower operational overhead, which is particularly beneficial in HAPS.

In [16], EH is studied on a dual-hop network where there is no direct communication between source and destination. In [17], exact and asymptotic closed-form outage probabilities (OPs) are obtained for underlay multi-hop EH cognitive relay networks under Rayleigh fading. Reference [18] investigates the outage behavior of a cooperative network aided by an EH relay node in a slow fading scenario. Ghosh et al. [19] performed a comprehensive outage analysis of SWIPT-enabled bidirectional D2D communications under spectrum sharing conditions. Their work considers both DF and AF relaying, highlighting the performance gap in outage behavior and energy efficiency under varying fading parameters. Moreover, reference [20] addresses energy management issues in solar-powered wireless sensor networks using EH prediction techniques. The proposed modified energy technique improves prediction accuracy by optimizing the battery lifetime of sensor nodes.

In addition to the above-given EH literature, there a few number of studies that focus on the performance of VHetNet. Among them, [21] considers space–air–ground station integrated network (SAGIN), where they derive OP and error probability. In [22], the authors consider the communication between a satellite, HAPS, and a ground station, where the optimal HAPS is selected in the presence of scattering, route loss, and targeting errors. For this scenario, the authors derive OP by considering various impairments. The study conducted by Yahia et al. examines the performance of multicast networks utilizing a multi-hop communication approach where error probability and OP are analytically formulated [23]. In [24], the authors consider satellite communication system where a HAPS node is used as a relay in the downlink transmission. In this scenario, both the satellite and the HAPS are assumed to utilize solar energy for transmission power. The authors in [25] consider a VHetNet model that incorporates the concept of SAGIN communication to optimize the overall sum rate of users. Reference [26] uses interference alignment with a multi-antenna tethered balloon relay to maximize sum-rate in HAPS systems without requiring channel state information. Reference [27] examines the potential benefits of optimizing user association and beamforming strategies in a hybrid satellite-HAPS-ground network to maximize network-wide data throughput. In addition, reference [28] proposes an adaptive channel division media access control protocol that optimizes channel resources in highly dynamic UAV networks by adaptively adjusting control and service channel intervals. In parallel with these research-oriented contributions, international standards bodies and industrial consortia have begun to formalise HAPS technology. Recent developments in standardization and industry initiatives, such as 3GPP Release 17 for Non-Terrestrial Networks (NTN), ITU-R's recommendations for HAPS, and ESA's NAVISP programme—including the NAVHAPS, HAPS-PNT case, highlight the growing interest and feasibility of HAPS-to-HAPS based architectures in real-world wireless communication deployments [29]–[31].

**Motivation and Contributions:** It is not difficult to find that none of the existing works in the literature have addressed the integration of EH in the design of VHetNet to the best of our knowledge. To remedy the problem of energy constraints in HAPS systems, while providing alternative energy methods, we propose a novel EH model using TS protocol for HAPS systems. More precisely, we consider a HAPS network consisting of two HAPS systems and a ground unit. In this particular system, we assume that the first HAPS is in idle mode and functions as an energy source, serving as a feeder HAPS. Meanwhile, the second HAPS operates under energy constraints, harnessing RF energy from the first HAPS and utilizing it for communication with the ground unit. To quantify the overall performance of the proposed system, we derive closed-form OP, ergodic capacity (EC), and throughput expressions by considering Nakagami-$m$ and shadowed-Rician channel models. To quantify the overall performance of the proposed system, we first derive closed-form expressions for the outage probability (OP), ergodic capacity (EC), and throughput under both Nakagami-$m$ and shadowed-Rician fading. Moreover, we rigorously discuss the EH requirements, specifying how the harvested-power budget varies with transmit power, antenna gain, and zenith angle. All analytical results are corroborated by extensive Monte-Carlo simulations, which confirm their accuracy over the full range of operating conditions, including various shadowing parameters and attenuation levels.

This paper is organized as follows. Section II describes the signal and system model. Section III provides the performance analysis. Section IV examines the energy-harvesting requirements for HAPS systems, and Section V presents the numerical results and related discussion. Finally, Section VI concludes the paper.

A note on mathematical notations: $\mathbb{E}[.]$ and $\overline{(.)}$ indicate the expectation operator, $F_X(.)$ and $f_X(.)$ represent cumulative distribution function (CDF) and probability density function (PDF) of $X$ respectively, and $|.|$ denotes for the absolute value. For clarity and ease of reference, Table 1 compiles the definitions and numerical values of all parameters employed throughout the paper.

## II. SYSTEM AND CHANNEL MODEL

In the proposed system, we examine an architecture consisting of two HAPS systems and a ground unit. The first HAPS, $H_1$, is assumed to loiter idly in the stratosphere and therefore operates without energy constraints. By contrast, the lower platform $H_2$, located at 20 km, is susceptible to energy shortages because of volcanic-aerosol veils, and polar stratospheric clouds. Both HAPS systems are equipped with context-aware sensors that continuously monitor weather conditions and their respective energy levels. At the initial time $t = 0$, $H_2$ is in a critical energy state, and starts harvesting energy from $H_1$ using the TS protocol. During the first $T_{\text{EH}}$ time-slot, $H_2$ gathers energy from $H_1$. In the subsequent $T_{\text{TX}}$ time-slot, $H_2$ utilizes the harvested energy to transmit data to the ground station ($G$). This cycle ensures that $H_2$ can maintain its operations despite its energy constraints, with the total transmission time amounting to $T = T_{\text{TX}} + T_{\text{EH}}$ as depicted in Fig. 1.

In this model, the EH channel $g_1$ between $H_1$ and $H_2$ is modeled as Nakagami-$m$ distribution as there is line-of-sight (LOS) connectivity between $H_1$ and $H_2$ [21]. The transmission channel ($g_2$) between $H_2$ and $G$ follows Shadowed-Rician channel due to shadowing effects experienced by $G$ at the ground level based on the tall buildings located at the city center [32]. In the first hop, the harvested energy at $H_2$ can be written as

$$E_{H_2} = \eta P_{H_1} G_T^{H_1} G_R^{H_2} \mathcal{G}_{H_1 H_2} |g_1|^2 T_{\text{EH}}, \quad (1)$$

where $0 < \eta \leq 1$ is the energy conversion efficiency, $P_{H_1}$ is the transmit power of $H_1$, $g_1$ represents the fading coefficient between the $H_1$ and $H_2$, $T_{\text{EH}}$ indicates the EH slot, $\mathcal{G}_{H_1 H_2} =$ $10^{-L_{H_1 H_2}/10}$ represents the per-link linear gain between $H_1$ and $H_2$. Transmit and receive antenna gains between $H_1$ and $H_2$ are shown as $G_T^{H_1}$, and $G_R^{H_2}$, respectively. Finally, $L_{H_1 H_2}$ shows the RF path loss (in dB) which can be obtained as [21]

$$L_{PL}[dB] = L_F + L_R + L_A + L_O, \quad (2)$$

where $PL \in \{H_1 H_2; H_2 G\}$, $L_F$ indicates the free space path loss (in dB) which is written as $L_F = 20 \log\left(\frac{4 \pi d_{H_1 H_2}}{\lambda_r}\right)$ for $H_1$ to $H_2$ communication, and $L_F = 92.45 + 20\log(f_r) + 20\log(d_{H_2G})$ for $H_2$ to $G$ communication. In the above formulation, $f_r$ represents the frequency in GHz, $\lambda_r$ is the RF wavelength, $d_{H_1H_2}$ and $d_{H_2G}$ represents the propagation distances from $H_1$ to $H_2$ and $H_2$ to $G$ respectively, where $d_{H_1H_2} = (\mathcal{H}_{H_2} - \mathcal{H}_{H_1})\sec(\zeta_{H_1H_2})$ and $d_{H_2G} = (\mathcal{H}_{H_2} - \mathcal{H}_G)\sec(\zeta_{H_2G})$. Here, $\mathcal{H}_{H_1}$, $\mathcal{H}_{H_2}$, and $\mathcal{H}_G$ represent the altitudes of $H_1$, $H_2$ and $G$ respectively. Moreover, $\zeta_{H_1H_2}$ and $\zeta_{H_2G}$ are the zenith angles between $H_1$ to $H_2$, and $H_2$ to $G$ respectively. Moreover, $L_R$ indicates the attenuation due to rain in dB/km, $L_A$ is the gaseous atmosphere loss that can be calculated in [21] and $L_O$ stands for other fading or miscellaneous losses in dB, which may be caused by a number of factors, including incorrect antenna pointing, polarization mismatch, or antenna deterioration [21]. The rain attenuation $L_R$ is calculated using $\gamma_R$ as

$$\gamma_R = k(R_{0.01})^\alpha, \quad (3)$$

where $k$ and $\alpha$ are frequency-dependent coefficients and $R_{0.01}$ is the rain rate (mm/h) exceeded for 0.01% of an average year. Rain attenuation can be calculated as described in [33]:

$$L_R = \gamma_R L_E, \quad (4)$$

where $L_E$ is the effective path length.

TABLE I
LIST OF NOTATIONS AND PARAMETERS.

| Parameter | Definition |
|---|---|
| $E_{H_2}$ | Harvested energy at $H_2$ |
| $T_{\text{TX}}$ | Data transmission slot |
| $T_{\text{EH}}$ | Energy harvesting slot |
| $\rho$ | Time-split ratio |
| $\gamma_0$ | instantaneous e2e SNR |
| $\overline{\gamma}_0$ | Average e2e SNR |
| $\eta$ | Energy conversion efficiency |
| $L_{H_1 H_2}$ | Path loss between the $H_1$ and $H_2$ (dB) |
| $L_{H_2 G}$ | Path loss between the $H_2$ and $G$ (dB) |
| $\mathcal{G}_{H_1 H_2}$ | Per-link gain between $H_1$ and $H_2$ |
| $\mathcal{G}_{H_2 G}$ | Per-link gain between $H_2$ and $G$ |
| $\mathcal{H}_{H_1}$ | Altitude of HAPS $H_1$ |
| $\mathcal{H}_{H_2}$ | Altitude of HAPS $H_2$ |
| $\mathcal{H}_G$ | Altitude of the ground station |
| $L_F$ | Free space path loss (dB) |
| $\lambda_r$ | RF wavelength (m) |
| $f_r$ | RF frequency (GHz) |
| $d_{H_1 H_2}$ | Propagation distance from $H_1$ to $H_2$ |
| $d_{H_2 G}$ | Propagation distance from $H_2$ to $G$ |
| $G_T^{H_1}$ | Transmit antenna gain of $H_1$ |
| $G_T^{H_2}$ | Transmit antenna gain of $H_2$ |
| $G_R^{H_2}$ | Receiver antenna gain of $H_2$ |
| $G_R^G$ | Receiver antenna gain of $G$ |
| $L_F$ | Free space path loss (dB) |
| $L_R$ | Attenuation due to rain (dB/km) |
| $L_A$ | Gaseous atmosphere loss (dB) |
| $L_O$ | Other fading or miscellaneous loss (dB) |
| $N_0$ | Noise power spectral density (dB) |
| $K_{dB}$ | Boltzmann's constant (dBW/K/Hz) |
| $T_N$ | Temperature of noise (dBK) |
| $B$ | Noise bandwidth (dBHz) |
| $m_1$ | Severity parameter of Nakagami-m channel |
| $m_2$ | Severity parameter of Shadowed-Rician channel |
| $\sigma_{g_1}^2$ | Variance |
| $\alpha, \beta$ | Shape parameters of Shadowed-Rician fading |
| $\Omega$ | Average power of LOS component |
| $b$ | Multipath component |
| $\gamma_{th}$ | Predetermined threshold value |
| $R_b$ | Number of bits transmitted per channel use (bpcu) |
| $\gamma_R$ | Specific attenuation (dB/km) |
| $R_{0.01}$ | Rain rate (mm/h) |
| $\zeta_{H_1 H_2}$ | Zenith angle between $H_1$ and $H_2$ (degree) |
| $\zeta_{H_2 G}$ | Zenith angle between $H_1$ and $G$ (degree) |
| $\mathcal{T}$ | Throughput (bpcu) |

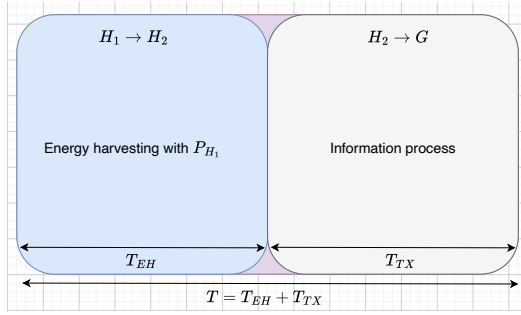

Fig. 1. Time schedule for EH and information process

In the second hop, $H_2$ uses the harvested energy to transmit its information to $G$. Consequently, during the second time interval of duration $T_{\text{TX}}$, the received signal at $G$ can be expressed as

$$y_G = \sqrt{P_{H_2} G_T^{H_2} G_R^G \mathcal{G}_{H_2G}}\, g_2 s + n_G, \quad (5)$$

where $s$ is the transmitted symbol with unit energy, $P_{H_2} = E_{H_2}/T_{\text{TX}} = \eta P_{H_1} G_T^{H_1} G_R^{H_2} \mathcal{G}_{H_1H_2} |g_1|^2 \left(\frac{\rho}{1-\rho}\right)$ is the transmit power of $H_2$, $T_{\text{TX}}$ denotes the data-transmission slot, the time-split ratio $\rho$ is given as $\rho = T_{\text{EH}}/(T_{\text{EH}} + T_{\text{TX}})$, per-link linear gain between $H_2$ and $G$ is given as $\mathcal{G}_{H_2G} = 10^{-L_{H_2G}/10}$, transmit and receive antenna gains between $H_2$ and $H_G$ are written as $G_T^{H_2}$, and $G_R^G$, respectively. Finally, $n_G \sim \mathcal{CN}(0, N_0)$ is the zero-mean complex Gaussian noise with $N_0$ power spectral density observed at $G$. With the aid of (1), (2) and (5), the instantaneous signal-to-noise ratio (SNR) at $G$ ($\gamma_0$) can be expressed as

$$\gamma_0 = \left(\frac{\eta\rho}{1-\rho}\right)\frac{P_{H_1}}{N_0}\,\mathcal{G}_{H_1H_2}\,\mathcal{G}_{H_2G}\,|g_1|^2|g_2|^2, \quad (6)$$

where $N_0$ is the single–sided noise power spectral density, expressed in linear units as $N_0 = k_B T_N B$ or, equivalently, in decibels as $N_0\,[\text{dB}] = K_{\text{dB}} + T_N\,[\text{dBK}] + B\,[\text{dBHz}]$ with $K_{\text{dB}} = -228.6$ dBW/K/Hz. The average SNR, i.e. the expectation of (6) over the unit–mean fading coefficients $|g_1|^2$ and $|g_2|^2$, is

$$\overline{\gamma}_0 \triangleq \mathbb{E}[\gamma_0] = \left(\frac{\eta\rho}{1-\rho}\right)\frac{P_{H_1}}{N_0}\,G_T^{H_2}G_R^{H_1}G_T^{H_1}G_R^G \mathcal{G}_{H_1H_2}\,\mathcal{G}_{H_2G}, \quad (7)$$

so that the instantaneous SNR can be written in the standard normalized form $\gamma_0 = \overline{\gamma}_0\,|g_1|^2|g_2|^2$.

## III. PERFORMANCE ANALYSIS
### A. Statistical Properties of Fading Channels

Here, we first obtain the statistical properties of fading channels including PDF and CDF of $\gamma_0$. To do that, we assign the random variables as $X = |g_1|^2$, $Y = |g_2|^2$, and $Z = |g_1|^2|g_2|^2 = XY$. Therefore, the CDF of $Z$ can be obtained as

$$F_Z(z) = \int_0^\infty f_X(x) \int_{-\infty}^{\frac{z}{x}} f_Y(y)\,dy\,dx \\ + \int_{-\infty}^0 f_X(x) \int_{\frac{z}{x}}^\infty f_Y(y)\,dy\,dx. \quad (8)$$

As $H_1$ to $H_2$ communication experiences Nakagami-$m$ distribution, the PDF of $X$ is given as

$$f_X(x) = \frac{x^{m_1-1}\exp\left(\frac{-x}{2\sigma_g^2}\right)}{\Gamma(m_1)(2\sigma_g^2)^{m_1}}, \quad (9)$$

where $m_1$ stands for the severity parameter of Nakagami-m fading, and $\sigma_{g_1}^2$ is the variance. In the second phase, $H_2$ to $G$ communication experiences shadowed-Rician fading and the PDF of $Y$ can be written as

$$f_Y(y) = \alpha\exp(-\beta y)\,_1F_1(m_2;1;\delta y), \quad (10)$$

where $m_2$ is the severity parameter of Shadowed-Rician fading channel, $\alpha = \frac{1}{2b}\left(\frac{2bm_2}{2bm_2+\Omega}\right)^{m_2}$, $\beta = \frac{1}{2b}$ and $\delta = \frac{\Omega}{2b(2bm_2+\Omega)}$. Here, $_1F_1(\cdot;\cdot;\cdot)$ is the confluent hypergeometric function of the first kind and the average power of the LOS component is represented by the parameter $\Omega$, whereas the multipath component is represented by $b$ [34]. Furthermore, for mathematical brevity, we can express $_1F_1(m_2;1;\delta)$ as [34]

$$_1F_1(m_2;1;\delta) = \exp(\delta)\sum_{k=0}^{m_2}\frac{(1-m_2)_k(-\delta y)^k}{(k!)^2}, \quad (11)$$

where $(\cdot)_k$ is the Pochhammer symbol [23]. Finally, by substituting (9), (10) and (11) into (8), we can express $F_Z(z)$ as

$$F_Z(z) = \frac{\alpha}{\Gamma(m_1)(2\sigma_{g_1}^2)^{m_1}}\sum_{k=0}^{m_2}\frac{(1-m_2)_k(-\delta)^k}{(k!)^2} \\ \times \int_0^\infty \int_0^{\frac{z}{x}} x^{m_1-1} y^k \exp\left(\frac{-x}{2\sigma_{g_1}^2} - y(\beta-\delta)\right)dy\,dx. \quad (12)$$

## B. Outage Probability

The OP can be described as the probability of instantaneous SNR ($\gamma_0$) falls below a predetermined threshold $\gamma_{th}$. Mathematically, OP can be expressed as [22]

$$P_{\text{out}} = \Pr[\gamma_0 \leq \gamma_{th}] = F_{\gamma_0}(\gamma_{th}), \quad (13)$$

where $F_{\gamma_0}(\gamma_{th})$ is the CDF of the e2e SNR at $G$ and we determine the threshold value as $\gamma_{th} = 2^{2R_b} - 1$ where $R_b$ is the number of bits transmitted per channel usage (bpcu). By inserting (4) into (11), we can obtain $F_{\gamma_0}(\gamma_{th})$ as

$$F_{\gamma_0}(\gamma_{th}) = \Pr\left(\overline{\gamma}_0 Z \leq \gamma_{th}\right) = F_Z\left(\frac{\gamma_{th}}{\overline{\gamma}_0}\right), \quad (14)$$

and then by substituting (14) in (12), we can express OP as can be seen in (15) at the top of the next page. By utilizing [35, eqn.(3.351.1)] in (15), $F_{\gamma_0}(\gamma_{th})$ can be written more tractable as

$$F_{\gamma_0}(\gamma_{th}) = \frac{\alpha}{\Gamma(m_1)(2\sigma_{g_1}^2)^{m_1}} \sum_{k=0}^{m_2} \frac{(1-m_2)_k(-\delta)^k}{k!(\beta-\delta)^{k+1}}$$
$$\times \int_0^\infty x^{m_1-1} \exp\left(\frac{-x}{2\sigma_{g_1}^2}\right) dx - \frac{\alpha}{\Gamma(m_1)(2\sigma_{g_1}^2)^{m_1}}$$
$$\times \sum_{k=0}^{m_2} \sum_{i=0}^{k} \frac{(1-m_2)_k(-\delta)^k}{i!k!(\beta-\delta)^{k-i+1}} \left(\frac{\gamma_{th}}{\overline{\gamma}_0}\right)^i$$
$$\times \int_0^\infty x^{m_1-1-i} \exp\left(\frac{-x}{2\sigma_{g_1}^2} - \frac{\gamma_{th}(\beta-\delta)}{\overline{\gamma}_0 x}\right) dx. \quad (16)$$

By using [35, eqn.(3.351.3)] in (14) and with few manipulations, OP can be expressed as

$$P_{\text{out}} = \alpha \sum_{k=0}^{m_2} \frac{(1-m_2)_k(-\delta)^k}{k!(\beta-\delta)^{k+1}} - \frac{2\alpha}{\Gamma(m_1)(2\sigma_{g_1}^2)^{m_1}}$$
$$\times \sum_{k=0}^{m_2} \sum_{i=0}^{k} \frac{(1-m_2)_k(-\delta)^k}{i!k!} \left(\frac{\gamma_{th}}{\overline{\gamma}_0}\right)^{\frac{m_1+i}{2}}$$
$$\times (\beta-\delta)^{\frac{m_1+i-2k-2}{2}} (2\sigma_{g_1}^2)^{\frac{m_1-i}{2}}$$
$$\times K_{m_1-i}\left(2\sqrt{\frac{\gamma_{th}(\beta-\delta)}{\overline{\gamma}_0 2\sigma_{g_1}^2}}\right), \quad (17)$$

where $K_{m_1-i}(\cdot)$ is the $m_1 - i$-th modified Bessel function of the second type [36].

## C. Ergodic Capacity

EC ($\mathcal{C}_e$) shows the highest rate of information transfer over a wireless channel and it is expressed in bpcu, as [23]

$$\mathcal{C}_e = \mathbb{E}[\log_2(1+\gamma_0)], \quad (18)$$

The closed-form expression of EC cannot be derived by using (16). However, it can be precisely upper-bound by using Jensen's inequality as

$$\mathcal{C}_e^{up} = \log_2(1+\mathbb{E}[\gamma_0]), \quad (19)$$

and $\mathbb{E}[\gamma_0]$ can be obtained as

$$\mathbb{E}[\gamma_0] = \int_0^\infty \gamma_0 f_{\gamma_0}(\gamma_0)\, d\gamma_0. \quad (20)$$

Herein, $f_{\gamma_0}(\gamma_0)$ shows the PDF of $\gamma_0$, and it can be found by taking the derivative of $F_{\gamma_0}(\gamma_0)$ as

$$f_{\gamma_0}(\gamma_0) = \frac{m_1^{m_1}\alpha}{\overline{\gamma}_0 \Gamma(m_1)\Omega^{m_1}} \sum_{k=0}^{m_2} \frac{(1-m_2)_k(-\delta\gamma_0)^k}{(k!)^2(\overline{\gamma}_0)^k}$$
$$\times \int_0^\infty y^{m_1-k-2} \exp\left(-\frac{ym_1}{\Omega} - \frac{\beta\gamma_0}{\overline{\gamma}_0 y}\right) dy. \quad (21)$$

Thereafter, by substituting (19) into (18), we can obtain $\mathbb{E}[\gamma_0]$ as

$$\mathbb{E}[\gamma_0] = \frac{m_1^{m_1}\alpha}{\overline{\gamma}_0 \Gamma(m_1)\Omega^{m_1}} \sum_{k=0}^{m_2} \frac{(1-m_2)_k(-\delta)^k}{(k!)^2(\overline{\gamma}_0)^k}$$
$$\times \int_0^\infty \int_0^\infty y^{m_1-k-2} \gamma_0^{k+1} \exp\left(-\frac{xm_1}{\Omega} - \frac{\gamma_0(\beta-\delta)}{\overline{\gamma}_0 y}\right) dy\, d\gamma_0. \quad (22)$$

By using [35, eqn.(3.351.3)] and with few manipulations, $\mathbb{E}[\gamma_0]$ can be expressed as

$$\mathbb{E}[\gamma_0] = \frac{m_1^{m_1}\alpha}{\overline{\gamma}_0 \Gamma(m_1)\Omega^{m_1}} \sum_{k=0}^{m_2} \frac{(1-m_2)_k(-\delta)^k(k+1)!}{(k!)^2(\overline{\gamma}_0)^k}$$
$$\times \left(\frac{\overline{\gamma}_0}{\beta-\delta}\right)^{k+2} \int_0^\infty x^{m_1} \exp\left(-\frac{m_1 x}{\Omega}\right) dx. \quad (23)$$

By applying [35, eqn.(3.351.3)] into (21), $\mathbb{E}[\gamma_0]$ can be written as

$$\mathbb{E}[\gamma_0] = \frac{\overline{\gamma}_0 \alpha \Omega}{(\beta-\delta)^2} \sum_{k=0}^{m_2} \frac{(1-m_2)_k(k+1)}{k!} \left(\frac{\delta}{\delta-\beta}\right)^k. \quad (24)$$

By substituting (22) into (17), we can obtain $\mathcal{C}_e^{up}$, as can be seen in (23) at the top of the next page.

## D. Throughput Analysis

Throughput refers to the rate at which data is successfully transmitted from one point to another over the network. It is expressed in bits per second, and from a mathematical perspective, it can be expressed as

$$\mathcal{T} = R_b(1 - F_{\gamma_0}(\gamma_{th})). \quad (26)$$

By substituting $R_b$ and $F_{\gamma_0}(\gamma_{th})$ into (24), $\mathcal{T}$ can be obtained.

## IV. ENERGY HARVESTING REQUIREMENT FOR HAPS SYSTEMS

Typical HAPS systems require power for propulsion, payload, and communications, and must ensure a received power level of approximately $-80$ dBm at the user to provide good/excellent connectivity [37]. Consequently, the link should tolerate large attenuations (on the order of 120–125 dB) depending on range, frequency, weather, and pointing. The total power budget can be written as

$$P_{\text{req}} = P_{\text{propulsion}} + P_{\text{payload}} + P_{\text{comm}}, \quad (27)$$

$$F_{\gamma_0}(\gamma_{th}) = \frac{\alpha}{\Gamma(m_1)(2\sigma_{g_1}^2)^{m_1}} \sum_{k=0}^{m_2} \frac{(1-m_2)_k(-\delta)^k}{(k!)^2} \int_0^\infty \int_0^{\frac{\gamma_{th}}{\overline{\gamma}_0 x}} x^{m_1-1} \exp\left(\frac{-x}{2\sigma_{g_1}^2} - y(\beta-\delta)\right) y^k \, dy \, dx. \quad (15)$$

$$\mathcal{C}_e^{up} = \log_2\left(1 + \frac{\overline{\gamma}_0 \alpha \Omega}{(\beta-\delta)^2} \sum_{k=0}^{m_2} \frac{(1-m_2)_k(k+1)}{k!}\left(\frac{\delta}{\delta-\beta}\right)^k\right). \quad (25)$$

where representative values (e.g., $P_{\text{propulsion}} \approx 100$ W and $P_{\text{payload}} \approx 40$ W) are adopted for illustration, while the communications power is determined by the link budget needed to achieve the $-80$ dBm target at $G$.

Using the proposed RF energy-harvesting (EH) model, the harvested power at $H_2$ (over the harvesting interval $T_{EH}$) is given by (1). To assess practical applicability, we perform a case-based evaluation that identifies parameter regimes ($\eta$, $\zeta_{H_1 H_2}$, $G_T^{H_1}$, $G_R^{H_2}$, path loss, and additional impairments) under which the harvested power meets or exceeds the communications power implied by the link budget at the operating frequency of 17.7 GHz. This framing avoids prescribing a fixed transmit power and instead ties feasibility to realistic antenna gains, propagation, and efficiency factors.

### A. Case I: High-Zenith Angle Alignment Between HAPS Nodes

In the first scenario, we consider a configuration where the maximum antenna gain achievable on the HAPS platforms is set to $G_T^{H_1} = G_R^{H_2} = 50$ dBi, representing a realistic upper bound for high-gain reflectarray or phased-array antennas deployable on HAPS systems. The zenith angle between the two HAPS nodes is $\zeta_{H_1 H_2} = 70°$, capturing a relatively wide separation geometry while maintaining line-of-sight alignment, and the platform altitudes are chosen as $\mathcal{H}_{H_1} = 20$ km and $\mathcal{H}_{H_2} = 21$ km, respectively—values consistent with typical stratospheric operations. The link operates in the Ka-band at $f = 17.7$ GHz with an efficiency factor of $\eta = 1$. .

Even when the zenith angle between $H1$ and $H_2$ increases to $70°$, the harvested power at $H_2$ remains at 26.2 dBm—a level that can easily satisfy the $-80$ dBm sensitivity threshold at the ground level [1], providing a very comfortable link margin. Moreover, because the first hop ($H_1 \to H_2$) retains ample link gain, this margin remains acceptable even if the $H_1$–$H_2$ separation is further increased. Hence, with $P_{H_1} = 200$ W and $\eta = 1$, efficient aerial-to-aerial power delivery and robust downlink communication can be sustained under large zenith angles, reinforcing the feasibility of the proposed high-altitude network architecture.

---

[1] Please note that the worst-case harvested-power level predicted herein can be achieved in practice with an $8 \times 10$ patch–rectifier array similar to the 10-W prototype demonstrated in [38], which supplies the required DC output while keeping every cell within its linear region and thus preserves the $-80$ dBm ground-station sensitivity margin. If such a patch array is not employed, the same margin can be maintained by selecting slightly lower transmit-power or antenna-gain settings so that a single rectifier operates at (or just below) its 16.6 dBm saturation level.

### B. Case II: Low-Zenith Angle Alignment Between HAPS Nodes

In the second scenario, we retain all system parameters from the previous case, except that the zenith angle between the HAPS nodes is reduced from $60°$ to $10°$, corresponding to a vertically aligned HAPS–to–HAPS link. This geometry minimizes the propagation distance and, consequently, the free-space path loss. With a reduced transmit power of $P_{H_1} = 20$ W and the same antenna gains $G_T^{H_1} = G_R^{H_2} = 50$ dBi at $f = 17.7$ GHz, the harvested power at node $H_2$ is calculated to be $P_{H_2} = 25$ dBm. Despite the tenfold reduction in $P_{H_1}$ compared to the first scenario, the downlink from $H_2$ to the ground still satisfies the $-80$ dBm sensitivity target, indicating a very comfortable link margin. Hence, vertical alignment not only boosts power-transfer efficiency but also drastically reduces the required transmit-power budget. These results confirm that efficient aerial-to-aerial energy transfer and robust downlink communication remain viable even under substantially lower transmit powers, offering a highly power-efficient baseline for sustainable high-altitude network architectures.

### C. Case III: Impact of Energy Conversion Efficiency

In this scenario, we retain the high-gain, near-vertical HAPS-to-HAPS configuration, fixing the antenna gains at $G_T^{H_1} = G_R^{H_2} = 50$ dBi and setting the zenith angle to $\zeta_{H_1 H_2} = 10°$. Even when the transmit power is reduced to $P_{H_1} = 20$ W, extremely small energy–conversion efficiencies or larger $H_1$–$H_2$ separations still yield a downlink from $H_2$ that comfortably satisfies the $-80$ dBm sensitivity requirement at the ground station, testifying to the generous link margin afforded by this geometry.

For reference, if the transmit power is restored to $P_{H_1} = 200$ W and the conversion efficiency falls to $\eta = 0.1$, the harvested power at $H_2$ nevertheless reaches 25 dBm. Compared with the high-zenith case, this underscores the robustness of near-vertical aerial links against energy-conversion inefficiencies, making them particularly attractive for reliable HAPS-to-HAPS energy transfer under practical hardware constraints. Moreover, because aerial-to-aerial power delivery at stratospheric altitudes is largely immune to atmospheric absorption, scattering, and obstruction, the effective conversion efficiency is expected to remain close to its ideal value ($\eta \approx 1$) in practice. Thus, even in the presence of sub-optimal hardware or unexpected losses, the harvested energy remains sufficient to meet the required power levels.

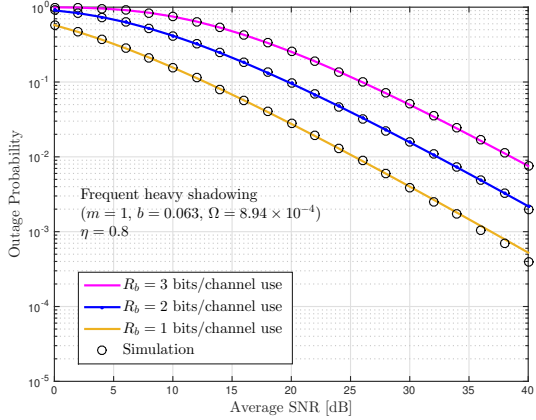

Fig. 2. OP versus average SNR for frequent heavy shadowing level and different $R_b$ values.

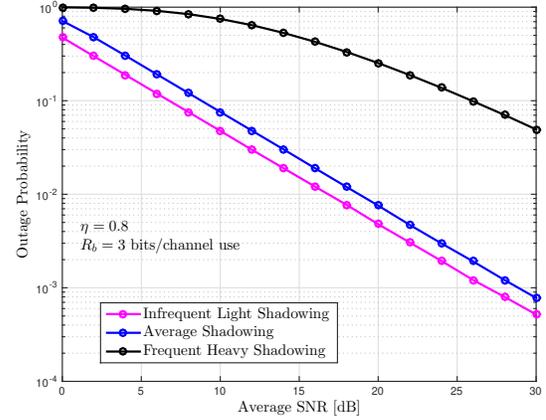

Fig. 4. OP versus average SNR for different shadowing levels.

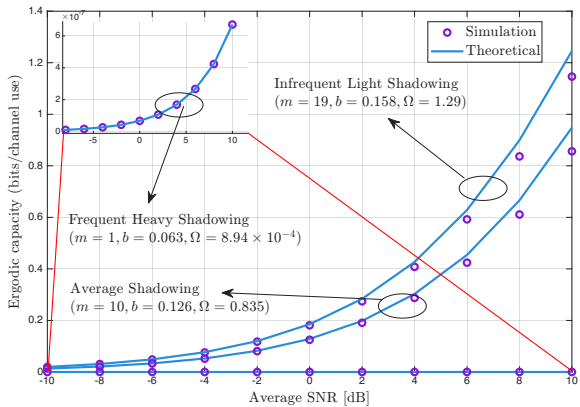

Fig. 3. EC versus average SNR in dB for different shadowing levels.

TABLE II
**SIMULATION PARAMETERS**

| Parameter | Value |
|---|---|
| $L_A$ between $H_1$ and $H_2$ | 0.0216 dB |
| $L_A$ between $H_2$ and $G$ | 0.0108 dB |
| $L_O$ | 5 dB |
| $\mathcal{H}_{H_1}$ | 21 km |
| $\mathcal{H}_{H_2}$ | 20 km |
| $\mathcal{H}_G$ | 0 km |
| $d_{H_1 H_2}$ | 1 km |
| $d_{H_2 G}$ | 20 km |
| $G_T^{H_1}$ | 50 dB |
| $G_R^{H_2}$ | 50 dB |
| $G_T^{H_2}$ | 52 dB |
| $G_R^G$ | 60 dB |
| $f_r$ | 17.7 GHz |
| $\zeta_{H_1 H_2}$ | 75° |
| $\zeta_{H_2 G}$ | 0° |
| $K_{\text{dB}}$ | $-228.6$ dBW/K/Hz |
| $T_N$ | 22.3805 dBK |
| $N_0$ | $-130.1995$ dB |
| $B$ | 76.02 dBHz |
| $\rho$ | 0.5 |

## V. NUMERICAL RESULTS AND DISCUSSIONS

In this section, the theoretical results are first verified through a set of simulations, in which $10^6$ Monte-Carlo trials are conducted. Then, both OP and EC performances are depicted by using three different shadowing models: frequent heavy shadowing ($m_1 = m_2 = 1, b = 0.063, \Omega = 8.94 \times 10^{-4}$), average shadowing ($m_1 = m_2 = 10, b = 0.126, \Omega = 0.835$) and infrequent light shadowing ($m_1 = m_2 = 19, b = 0.158, \Omega = 1.29$) [23]. In addition, impact of rain attenuation is illustrated, and finally, we provide important design guidelines that can be helpful for the reader.

Throughout the simulations we set $m_1 = m_2 \triangleq m$, time-split ratio is set to $\rho = 0.5$, and the altitudes of HAPS systems and ground station are taken as $\mathcal{H}_{H_1} = 21$, $\mathcal{H}_{H_2} = 20$, and $\mathcal{H}_G = 0$ above the mean sea level. Moreover, the zenith angles are chosen as $\zeta_{H_1 H_2} = 75°$, and $\zeta_{H_2 G} = 0°$. Moreover, rain attenuation is demonstrated in Section V-C, and all parameters employed in the numerical simulations are summarized in Table II.

### A. Verifications of the Theoretical Expressions

Fig. 2 illustrates the OP performance as a function of average SNR for different $R_b$ values under the frequent heavy shadowing. In this figure, the solid lines show the theoretical findings whereas simulations are depicted with marker symbols. The figure clearly shows that the theoretical findings match well with the MC simulations. Furthermore, we observe that the performance improves when the value of $R_b$ decreases as it is easy to transmit 1 bpcu rather than 3 bpcu.

Fig. 3 illustrates the EC as a function of the average SNR for all shadowing levels when $\eta = 0.2$ and $R_b = 3$ bpcu. At 10 dBm average SNR, the EC difference between frequent heavy shadowing and light shadowing is about 0.65 bit/s/Hz. We can observe from the figure that when shadowing level decreases, EC performance of the system enhances. Furthermore, the figure clearly shows that the theoretical findings agree with the MC simulations in all shadowing levels.

### B. Impact of Shadowing and Data Rate

Fig. 4 shows the OP as a function of average SNR for all shadowing models. We can observe from the figure that in frequent heavy shadowing, the OP achieves $10^{-1}$ performance at about 25 dB. Furthermore, we can realize that the proposed model under-performs at heavy shadowing whereas the best performance can be obtained in the presence of infrequent light shadowing.

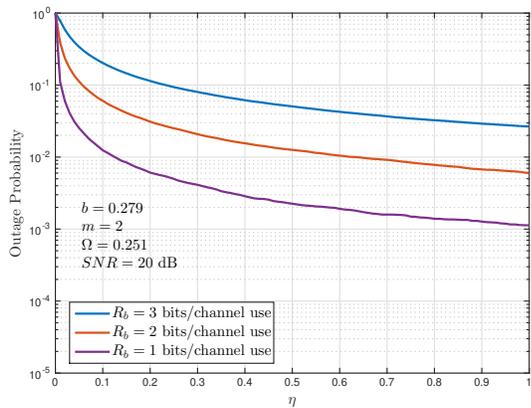

Fig. 5. OP versus $\eta$ for different $R_b$ values for SNR = 20 dB.

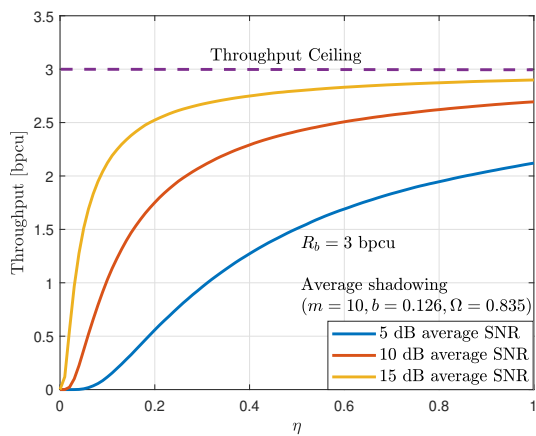

Fig. 6. Throughput versus average SNR for different $R_b$ values.

Fig. 5 illustrates the OP as a function of EH factor when $b = 0.279, m = 2, \Omega = 0.251$ and average SNR is 20 dB for different $R_b$ values. We observe from the figure that when $\eta \to 1$, OP performance improves due to better EH capability. Moreover, it is inferred from the figure that the best OP is observed at the $R_b = 1$ bpcu as in Fig. 4. As the $\eta$ value approaches 1, the OP curve flattens and converges to a fixed level. Moreover, when $\eta = 1$, the OP achieves $10^{-3}$ performance for $R_b = 1$ bpcu which can be considered as an acceptable outage rate at 20 dB.

Fig. 6 depicts the throughput of the proposed system with respect to average SNR for different $R_b$ levels. As observed from the figure, even in the ideal scenario where EH is perfect ($\eta = 1$), it is evident that achieving the ideal throughput ceiling (3 bpcu) is almost impossible under low SNR conditions in the presence of average shadowing effects. In addition, based on the EH capability, an SNR of either 5 dB or 10 dB would be sufficient to achieve a data rate of 2 bpcu.

### C. Impact of Rain Attenuation

By considering three representative rain-rate conditions—light rain ($R_{0.01} = 2$ mm/h), moderate rain ($R_{0.01} =$

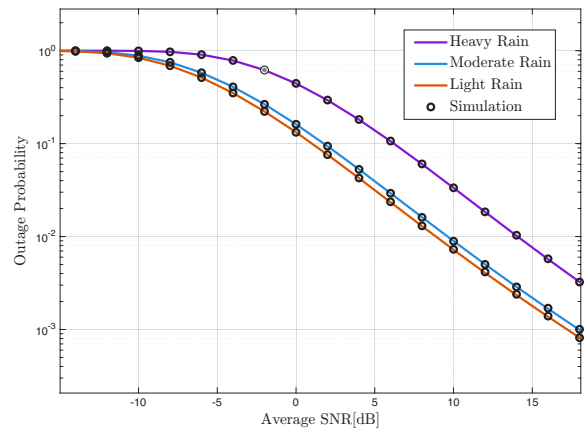

Fig. 7. OP versus average SNR for different rain levels.

10 mm/h), and heavy rain ($R_{0.01} = 50$ mm/h), a set of simulations has been performed at the Ka-band (17.7 GHz) to evaluate how rain attenuation impacts the link quality between the HAPS systems and the ground unit. These parameters capture the full range of practical rainfall conditions and allow the Ka-band performance to be assessed under realistic propagation scenarios.

Figure 7 illustrates the outage probability (OP) as a function of the average SNR for three different rain scenarios including light rain, moderate rain, and heavy rain. Both theoretical and simulation results are included in the figure to validate the accuracy of the analytical model. The curves indicate that as the rain intensity increases, the OP performance worsens, reflecting the significant impact of rain attenuation on system reliability. For light rain, the OP performs substantially better than heavy rain, highlighting the sensitivity of the system to varying rain conditions.

### D. Design Guidelines

In this section, we provide some crucial design recommendations that may be beneficial in the design of HAPS networks.

- The results show that with effective EH strategies, HAPS systems can sustain their operational efficiency and reliability, even under heavy rain, and heavy shadowing effects. This ability is especially important in situations where conventional power sources are scarce or inaccessible. Therefore, the successful integration of EH in HAPS systems could greatly improve their sustainability and overall performance, making them a dependable option for reliable communication networks.
- The throughput analysis reveals that the system's communication capabilities are severely diminished under frequent heavy shadowing conditions. Even at medium and high SNR levels, the number of symbols that can be transferred significantly decreases. As mentioned above, it is crucial to select a high-ground location for the ground station to avoid areas with tall buildings or dense urban environments. By doing so, reliable communication can be established.
- Energy-harvesting analyses in Section IV show that link viability is governed by three parameters: the combined

AUTHOR et al.: PREPARATION OF PAPERS FOR IEEE TRANSACTIONS AND JOURNALS (MAY 2024) 9

antenna gain, the zenith angle $\zeta_{H_1 H_2}$, and the RF-to-DC efficiency $\eta$. The results indicate that maintaining a 50 dBi gain per HAPS aperture and keeping the slant angle nearly vertical enable the system to meet the $-80$ dBm ground-station sensitivity target even when $\eta$ falls to 0.1 or $P_{H_1}$ is reduced to 20 W.

- The outage probability analysis, illustrated in Figure 7, demonstrates the significant impact of rain attenuation on system performance. As rain intensity increases from light to heavy rain, the outage probability deteriorates, highlighting the importance of adaptive power control strategies. For instance, in heavy rain scenarios, achieving acceptable communication quality may require substantial increases in transmission power or the deployment of additional resources.

- Based on the simulation results, it is recommended that ground stations be placed in areas with minimal obstructions, such as open rural or semi-urban regions, to avoid the severe degradation caused by intense shadowing effects. Moreover, positioning the HAPS system to optimize zenith angles can minimize path loss and improve overall system performance.

## VI. CONCLUSION

In this paper, we propose an EH strategy specifically designed for HAPS systems. This innovative approach allows a HAPS system facing energy constraints, to gather energy from another HAPS system that has no energy constraints. This strategy ensures the continuous operation of the energy-limited HAPS system by utilizing the harvested energy. To thoroughly evaluate the overall performance of the proposed system, we derived the OP and EC expressions and validated these metrics through MC simulations. The results indicate that with EH strategies, HAPS systems can maintain their operational efficiency and reliability. This capability is particularly crucial in scenarios where traditional power sources are scarce or unavailable. Consequently, the successful implementation of EH in HAPS systems could significantly enhance their sustainability and overall performance.